# Lattice of infinite bending-resistant fibers

V.Kobelev[1]

[1] kobelev@imr.mb.uni-siegen.de, University of Siegen, D-57076, Siegen, Germany

**Nomenclature**

| | |
|---|---|
| $w = e^{i\theta}, \quad z = e^{i\eta}$ | Complex units |
| $(u_{mn}, v_{mn})$ | Vector of displacements of fiber contact point with the number $(m, n)$ |
| $(r_{mn}, s_{mn})$ | reactions between fibers at contact point with the number $(m, n)$ |
| $(p_{mn}, q_{mn})$ | external applied fiber contact point with the number $(m, n)$ |
| $T(w) = \sum_{k=-\infty}^{\infty} w^k t_k$ | Green coefficients for fiber tension |
| $B_\lambda(z) = \sum_{k=-\infty}^{\infty} z^k b_k(\lambda)$ | Green coefficients for fiber bending |
| $\alpha = Ed^4/64$ | bending stiffness of fibers |
| $\beta = Ed^2/4$ | longitudinal stiffness of fibers |
| $\lambda$ | uniform tension force in X-fibers |
| $\gamma$ | uniform tension force in Y-fibers |
| $\eta = \Delta\sqrt{\lambda/\alpha}$ | dimensionless coefficient |
| $d$ | diameters of fibers |
| $\Delta$ | distances between fibers |
| $E$ | Young modulus of fibers |
| $f_n$ | external forces in Y-direction applied on the crack edges |
| $v_{n0}$ | defections of the crack edges in Y-direction |
| $T^*(\xi), B_\lambda^*(\xi)$ | Fourier transformations of the differential operators |
| $u(x, y), v(x, y)$ | components of macroscopic displacement |
| $p(x, y), q(x, y)$ | components of macroscopic surface load |
| $v(x, 0)$ | macroscopic crack opening |

Capital letters denote Fourier series of the corresponding infinite vectors. Minus in the subscript denotes principal, plus - regular part of Laurent expansions, e.g.

$$W_+(z) = \sum_{n=0}^{\infty} w_n z^n, \qquad W_-(z) = \sum_{n=-\infty}^{-1} w_n z^n, \qquad W(z) = \sum_{n=-\infty}^{\infty} w_n z^n.$$

# 1 The double-periodic 2D lattice of infinite fibers

## 1.1 *The modeling of cellular and lattice materials*

In modern engineering practice cellular materials find broad application because of their ultimate mechanical properties. The cellular materials provide outstanding bearing capacity in light weight structures. The physical properties of cellular materials such as good thermal insulation and energy absorption make these materials suitable for temperature and vibration reduction. The majority of bulk structural materials, such metals and concrete, fail predominantly under tensile loading. For the cellular materials the dangerous loading is compression, because the compression leads to compression instability and wrinkling. Consequently, the design of cellular materials requires the biaxial tension over the whole surface. In many design applications they are used under tensile loading. If one component of tension field vanishes, the material changes its mechanical behavior. As the biaxial tension field in cellular material degrades locally into the uniaxial tension field the elastic behavior of the material alters. The uniaxial tension may be considered as the transition case between two-axial tension and the state of mixed compression and tension. If the parent material is brittle the dominant failure mode in this condition is brittle fracture caused by a propagating crack.

## 1.2 *The methods of lattice mechanics*

There are two main approaches for calculation of fracture toughness of cellular materials. One approach is based of the lattice mechanics [1]. The majority of studies of cellular materials are based on an assumption, that cellular material is modeled by a periodic lattice composed of elements that connected at the nodal points. The elements are short rods or Euler beams [2].

The mechanics of failure for elastic-brittle lattice materials is reviewed in [3,4]. Closed-form expressions are summarized for fracture toughness as a function of physical parameters for a wide range of periodic lattices. The lattice or spring network models are discussed from the point of view for micromechanics applications. The models discussed in the review have their origin in the atomistic representations of matter on one hand, and in the truss-type systems in engineering on the other. The brittle fracture behavior of periodic 2D cellular material weakened by a system of non-interacting cracks is investigated in [5]. The material is represented as a lattice consisting of rigidly connected Euler beams which can fail when the skin stress approaches a given boundary value.

The paper [6] addresses the issue of creating a lattice model suitable for design purposes and capable of quantitative estimates of the mechanical properties of a disordered microstructure.

The imperfection sensitivity of in-plane modulus and fracture toughness was explored in [7]. The study was performed for several morphologies of 2D lattice: the isotropic triangular, hexagonal and Kagome lattices, and two orthotropic square lattices. The elastic lattices were assumed to fail when the maximum local tensile stress at any point attains the tensile strength of the solid. The assumed imperfection comprised a random dispersion of the joint position from that of the perfect lattice.

The microscopic modeling was performed in [8] to study the effects of several physical factors on dislocation emission in a model hexagonal lattice. The method of calculation is the lattice Green s-function method, together with several pair potentials with variable parameters.

The methods of lattice mechanics are used as well for modeling of elastic properties of carbon-based materials [9]. The review mechanical properties of carbon-based materials is in [10, 11, 12].

## 1.3 *The shear-lag method*

The second approach, known as shear-lag method, describes the infinite fibers arranged in one direction. The fibers withstand tension are fixed together by the matrix. The matrix material transfers only shear force between fibers. The shear-lag method was introduced by Hedgepeth [13, 14] to analyze the multifilament failure problems of laminated composites. The shear-lag approach is based upon easy assumptions and usually provides adequate physical insights of complicated problems.

The shear-lag model is suited for two-dimensional multifilament failure problems of unidirectional fiber composites. The method allows obtaining an approximate analytical expression for the fracture toughness. The technique also has been extended to include the effects of plastic flow of the matrix and the condition of debonding between interfaces. The similar models of two-dimensional composite were developed in recent papers. The paper [15] examines the stress concentration problems in unidirectional continuous fiber composites due to transverse multifilament failure. Mechanics of fracture for composite materials and micromechanical mechanisms are summarized in Liu [16].

## 1.4  Assumptions of the model

In this section we examine the two-directional double-periodic fiber lattice composed of two families of infinite fibers.

The attribute of the model is the following. On one side, similarly to the cellular analysis the lattice consists of two fiber families, likewise to the cellular analysis. In contrast to the common lattice models the fibers in the actual model extend not only between the nodal points, but are of infinite length. The fibers of infinite length are known from shear-lag models. On the other side, in contrast to the shear-lag models the interaction between the fibers provides not the matrix, which is absent. The interaction between the fibers of one family offers namely the second family of orthogonal fibers.

Based on the developed model, the brittle fracture behavior of periodic 2D material weakened by a crack is investigated. The analysis considers a two-directional continuous fiber composite containing a slit notch in the transverse direction. The mode focuses specifically on the stress concentration factors of fibers adjacent to the cracks.

The objective of this paper is to give the mathematical description of fracture of a lattice which is contains a crack, and accurately calculate the stresses in fibers at crack tip vicinity. It is assumed, that the crack growth occurs, when the stress in the first unbroken fiber exceeds its maximal allowed value. The failure of the first fiber on the crack tip causes the stepwise elongation of the crack, such that the rupture of the first fiber determines the fracture toughness.

## 1.5  Modeling of double-periodic 2D lattice

In this section we derive the equations of double-periodic 2D lattice. Consider the infinite perfect lattice, composed of two sets of infinite fibers (**Fig.1**). The two-dimensional lattice is composed of two orthogonal rows of periodically spaced straight, bending-resistant fibers. The first family of fibers, which are parallel to X-axis, will be referred to X-fibers. The second family of fibers will be referred to Y-fiber. The distance between parallel fibers of each set is taken to be unit. The material is thus represented as a double-periodical lattice consisting of rigidly connected beams.

Both families of fibers are considered as axially preloaded elastic beams.

There are two possible systems of forces in the lattice. The first system is characterized as follows:

> *the Y-fibers withstand bending and transmit the load to the neighboring X-fibers and stretched in Y-direction;*
>
> *the X-fibers resist bending and transfer the load to the Y-fibers and stretched in X-direction.*

The constitutive equations for both systems of forces turn into each other after swapping X and Y.

Each X-fiber is pre-stressed with a constant tension force. The pretension in the X-fibers is assumed to be independent upon the deformation in Y-direction. Vice versa the pre-stress of Y-fibers is not influenced by the deformation of X-fibers and assumed to be given. These assumptions are physically valid for small deformations.

The first low index $m$ is used to numbering the X-fibers. The second index $k$ enumerates the Y-fiber. The pair $(m,k)$ marks the point where the X-fiber with number $m$ crosses the Y-fiber with the number $k$.

The external forces with components

$$(p_{mk}, q_{mk})$$

are applied at the cross-section points of the orthogonal fibers.

The fibers of both families are fastened together in the cross-section points. No relative displacements and no moments occur at contact point. The components of vector of reaction forces at the contact point $(m,k)$ are referred to as

$$(r_{mk}, s_{mk}).$$

The vector of nodal displacements the cross-section points at the contact point $(m,n)$ are

$$(v_{mn}, u_{mn}).$$

Both families of fibers behave linearly elastic. The Y-components of nodal displacements of are the linear functions of the Y-components nodal reactions and nodal applied forces:

(1.1)
$$v_{mn} = \sum_{k=-\infty}^{\infty} b_{k-n}(\lambda)(s_{mk} + q_{mk}),$$

$$v_{mn} = -\sum_{k=-\infty}^{\infty} t_{k-m} s_{kn}.$$

The displacements in X-direction depend on linearly nodal reactions and nodal applied forces in the X direction correspondingly

$$u_{mn} = \sum_{k=-\infty}^{\infty} b_{k-m}(\gamma)(r_{nk} + p_{nk}),$$

$$u_{mn} = -\sum_{k=-\infty}^{\infty} t_{k-n} r_{km}.$$

The infinite vectors

$$b_k \text{ and } t_k \quad (-\infty < k < \infty)$$

in the above equations represent the Green influence coefficients.

The Green coefficients $b_k$ determine the value of the displacement due to bending under the action of local force normal to fiber axis. The forces are applied in the point with the number $k$. These coefficients depend on the pre-tensions $\lambda$ or $\gamma$ in the directions of fiber axes. Thus, if the Y-fiber is considered, its pre-tension is $\gamma$. Correspondingly the pretension in X-fibers is $\lambda$. Mostly the latter case is used is this article as the crack along X-axis is considered.

The Green coefficients $t_k$ provide the axial displacement under the action of local force in the direction of fiber axis. The Green coefficients are determined in closed form in the next Section.

## 1.6  Solutions of finite-difference equations

The equations (1.1) are the finite-difference equations. For solution of the finite-difference equations the methods of the discrete Fourier transform are applied [17].

The unit complex numbers are introduced

$$w = e^{i\theta}, \quad z = e^{i\eta}, \quad |w| = 1, \quad |z| = 1.$$

The Fourier images of the displacements $V(z,w)$ and reaction forces $S(z,w)$ are the infinite sums

$$\begin{Bmatrix} S(z,w) \\ V(z,w) \end{Bmatrix} = \sum_{m=-\infty}^{\infty}\sum_{n=-\infty}^{\infty} w^n z^m \begin{Bmatrix} s_{mn} \\ v_{mn} \end{Bmatrix}.$$

Fourier images of the Green coefficients are

$$T(z) = \sum_{k=-\infty}^{\infty} z^k t_k ,$$

$$B_\lambda(w) = \sum_{k=-\infty}^{\infty} w^k b_k(\lambda).$$

The Fourier image of the initially known external forces $Q(z,w)$ reads

$$Q(z,w) = \sum_{m=-\infty}^{\infty}\sum_{n=-\infty}^{\infty} w^n z^m q_{mn} .$$

Performing discrete Fourier transform we reduce the equations (1.1) to the linear algebraic equations between the Fourier images

(1.2) $\quad V(z,w) = -T(w)S(z,w),$

$$V(z,w) = B_\lambda(w)[S(z,w) + Q(z,w)].$$

The first linear algebraic equation could be immediately solved with respect to $S(z,w)$. Elimination of nodal reactions $S(z,w)$ leads to the governing equations of Y-displacements in nodal points:

(1.3) $\quad V(z,w) = \dfrac{Q(z,w)}{T^{-1}(z) + B_\gamma^{-1}(w)}.$

The similar equation describes the X-displacements in nodal points

(1.4) $\quad U(z,w) = \dfrac{P(z,w)}{T^{-1}(w) + B_\gamma^{-1}(z)},$

with

$$\begin{Bmatrix} P(z,w) \\ U(z,w) \end{Bmatrix} = \sum_{m=-\infty}^{\infty}\sum_{n=-\infty}^{\infty} w^n z^m \begin{Bmatrix} p_{mn} \\ u_{mn} \end{Bmatrix}.$$

Important, that the stretching and bending of fibers in one direction are assumed to be small enough. The initial pre-stress is assumed much higher that the variation of force in the fibers due to elastic deformation.

That is why both equations (1.3) and (1.4) are independent and system obeys linear equations.

# 2  Green coefficients for periodically loaded infinite elastic fiber

## *2.1  Evaluation of Green coefficients for periodical loads*

In this Section the closed form expressions for the Green coefficients will be delivered.

The Green coefficients $b_k$ describe bending of infinite fibers, subjected to the periodically spaced local forces orthogonal to the direction of fiber.

The Green coefficients $t_k$ represent stretching of infinite fibers, subjected to the periodically spaced local forces in the direction of fiber.

Consider an infinite fiber subjected to the periodically spaced local forces of different magnitudes (**Fig.2**). If the direction of forces is parallel to the axis of fiber the stretching occurs. On the other hand, the forces with the orthogonal direction cause the bending of the fiber. In this section the relations between the values of concentrated forces and values of deflections of fiber in the force application points are derived.

The fibers are of an infinite length and the number of periodically spaced local forces is also infinite. The infinite sequence of forces can be characterized by means of their Fourier series. The coefficients of the Fourier series are the magnitudes of applied forces.

The deflections in the periodically spaced points are characterized similarly by their Fourier series.

Our goal is to find the relation between the Fourier series of applied loads and the Fourier series of displacements. We use for derivation one symbolic equation. This symbolic could be applied for both cases of bending and tension.

For this purpose the equations of the deformations of fibers are applied. The equations of both bending and tension of a fiber can be symbolically written in the form

(2.1)   $\mathrm{A}(\mathrm{D})v(x) = q(x), \quad -\infty < x < \infty.$

Here $\mathrm{D}$ is the symbolic operator of differentiation with respect to x

$$\mathrm{D} \equiv \frac{d}{dx}, \quad \mathrm{D}^2 \equiv \frac{d^2}{dx^2}, \quad \mathrm{D}^4 \equiv \frac{d^4}{dx^4},$$

$A(D)$      the polynomial differential operator;

$v(x)$      the continuous displacement and

$q(x)$      the continuous applied load in fiber direction or normal to fiber.

The integral Fourier transform is applied to the equation (2.1). The Fourier image of (2.1) is

(2.2)    $A(-i\xi)V^*(\xi) = Q^*(\xi)$,

where

$$\begin{Bmatrix} V^*(\xi) \\ Q^*(\xi) \end{Bmatrix} = \frac{1}{\sqrt{2\pi}} \int_{-\infty}^{\infty} \begin{Bmatrix} v(x) \\ q(x) \end{Bmatrix} e^{i\xi x} dx.$$

The Eq. (2.2) can be rewritten in the form

$$V^*(\xi) = R^*(\xi) Q^*(\xi),$$

where

$$R^*(\xi) = \frac{1}{A(-i\xi)}$$

is the resolvent for the differential operator (2.2).

The applied load is the periodically spaced concentrated forces. The distance between force application points is equal to $\Delta$. The applied load can be represented as the function with Dirac delta-functions [18]

$$q(x) = \sum_{n=-\infty}^{\infty} q_n \delta(x - n\Delta),$$

where

$q_n$ - the amplitudes of periodically spaced forces,

$\delta(x-y)$ - Dirac delta-function with the support in y point.

Fourier transform of the sequence $p_n$ reads

$$Q(w) = \sum_{n=-\infty}^{\infty} e^{in\theta} q_n,$$

(2.3) $$Q^*(\xi) = \frac{1}{\sqrt{2\pi}} \int_{-\infty}^{\infty} q(x) \exp(i\xi x) dx =$$

$$= \frac{1}{\sqrt{2\pi}} \int_{-\infty}^{\infty} \sum_{n=-\infty}^{\infty} q_n \delta(x - n\Delta) \exp(i\xi x) dx = \frac{1}{\sqrt{2\pi}} \sum_{n=-\infty}^{\infty} q_n \exp(i\xi n\Delta).$$

Consequently,

$$Q^*\left(\frac{\theta}{\Delta}\right) = \frac{1}{\sqrt{2\pi}} \sum_{n=-\infty}^{\infty} q_n \exp(i\theta n),$$

such that

(2.4) $$Q^*\left(\frac{\theta}{\Delta}\right) = \frac{1}{\sqrt{2\pi}} Q(w), \qquad w = e^{i\theta}.$$

The deflections at the stress application points are equal to

$$v_n \equiv v(n\Delta) = \frac{1}{\sqrt{2\pi}} \int_{-\infty}^{\infty} V^*(\xi) \exp(-in\xi\Delta) d\xi.$$

Fourier series of the sequence $v_n$ can be represented in terms of Fourier transforms. The Fourier transform for $v_n$ is

(2.5) $$V(w) = \frac{1}{\sqrt{2\pi}} \int_{-\infty}^{\infty} V^*(\xi) \exp(-in\xi\Delta) \sum_{n=-\infty}^{\infty} \exp in\theta dx =$$

$$= \frac{1}{\sqrt{2\pi}} \int_{-\infty}^{\infty} V^*(\xi) \exp(-in\xi\Delta) \left[2\pi \sum_{n=-\infty}^{\infty} \delta(\theta - 2\pi n)\right] dx =$$

$$= \frac{\sqrt{2\pi}}{\Delta_x} \sum_{n=-\infty}^{\infty} V^*\left(\frac{\theta - 2\pi n}{\Delta}\right).$$

The formula [4]

$$\sum_{n=-\infty}^{\infty} \exp(in\theta) = 2\pi \sum_{n=-\infty}^{\infty} \delta(\theta - 2\pi n).$$

is used in (2.5).

To derive the desired relation between $V(w)$ and $Q(w)$ Eq. (2.4) and (2.5) are used. With the function $R(w)$ the result $V(w)$ of action $Q(w)$ reads

(2.6) $\quad V(w) = R(w) Q(w),$

$$R(w) = \frac{1}{\Delta} \sum_{n=-\infty}^{\infty} R^* \left( \frac{\theta - 2\pi n}{\Delta} \right), \qquad w = e^{i\theta}.$$

The function $R(w)$ is the Fourier transform of the resolvent for the differential operator (2.2).

## 2.2 Green coefficients for bending of infinite fibers

Consider now the case of bending of an infinite fiber. The Bernoulli-Euler beam equation reads

(2.7) $\quad \alpha \dfrac{d^4 v}{dx^4} - \lambda \dfrac{d^2 v}{dx^2} = q \quad$ or $\quad \alpha D^4 v - \lambda D^2 v = q.$

Here $\quad v$ is the displacement normal to fiber axis due to bending,

$\quad q$ is the force in the direction normal to fiber axis,

$\quad \alpha$ is the bending stiffness,

$\quad \lambda$ is the pre-stress in fibers.

The corresponding expression for Fourier transform of the Green function is

(2.8) $\quad R^*(\xi) = \dfrac{1}{\alpha \xi^4 + \lambda \xi^2}.$

Using the formulas [19]

$$\sum_{n=-\infty}^{\infty} \frac{1}{(\theta - 2\pi n)^4} = \frac{3 - 2\sin^2(\theta/2)}{48 \sin^4(\theta/2)},$$

$$\sum_{n=-\infty}^{\infty} \frac{1}{(\theta - 2\pi n)^2 + s^2} = \frac{\sinh s}{2s(\cosh s - \cos \theta)},$$

the Fourier transform for the Bernoulli-Euler beam equation of the Green function reads

(2.9) $\quad R(w) \to B_\lambda(w) = \begin{cases} \dfrac{\Delta^3 \left[ 3 - 2\sin^2\left(\dfrac{\theta}{2}\right) \right]}{48 \alpha \sin^4\left(\dfrac{\theta}{2}\right)} & \text{for} \quad \lambda = 0, \\[2em] -\dfrac{2\sqrt{\alpha} \sinh(\eta)}{4\lambda^{3/2} \left[ \cosh(\eta) - 1 + 2\sin^2\left(\dfrac{\theta}{2}\right) \right]} + \dfrac{\Delta}{4\lambda \sin^2\left(\dfrac{\theta}{2}\right)} & \text{for} \quad \lambda > 0, \end{cases}$

where

$$\eta = \Delta\sqrt{\dfrac{\lambda}{\alpha}}$$

The behavior of material depends upon the presence of pre- tension in the orthogonal family of fibers.

## 2.3 Green coefficients for tension of infinite fibers

According to this model the equation of fiber is the following

(2.10) $\quad \beta \dfrac{d^2 v}{dx^2} = q,$

where $\beta$ is longitudinal stiffness of fibers,

$v$ is the displacement along fiber axis due to tension,

$q$ is the force in the direction of fiber axis,.

The Eq. (2.13) in operator notation reads

(2.11) $\quad \beta \mathrm{D}^2 v = q.$

From (2.3) we obtain

(2.12) $\quad \mathrm{R}^*(\xi) = \dfrac{1}{\beta \xi^2}.$

For the evaluation of infinite sum (2.6) the formula is applied

$$\sum_{n=-\infty}^{\infty} \dfrac{1}{(\theta - 2\pi n)^2} = \dfrac{1}{4\sin^2(\theta/2)}.$$

The Fourier transform for the Green coefficients (2.14) reads

(2.13)
$$R(w) \to T(w) = \dfrac{1}{\Delta}\sum_{n=-\infty}^{\infty}\dfrac{1}{\beta}\left(\dfrac{\Delta}{\theta - 2\pi n}\right)^2 =$$
$$= \dfrac{\Delta}{\beta(1-w)\left(1-\dfrac{1}{w}\right)} = \dfrac{\Delta}{4\beta \sin^2\left(\dfrac{\theta}{2}\right)}.$$

# 3 Lattice with semi-infinite crack

## *3.1 Loads and displacements on the crack edges*

The material is represented as a lattice consisting of rigidly connected Euler beams which can fail when the stress in fiber approaches an ultimate value. The fracture toughness is calculated and its dependence upon material parameters is examined. For this purpose the problem of a sufficiently long length crack in an infinite lattice produced by several broken beams is considered. The crack is assumed to by long enough to be considered as the semi-infinite. This assumption provides the mathematical advantage. Namely, the mathematical treatment of the lattice, that contains a semi-infinite crack, is simpler, that for a finite crack. Consequently, the lattice problem is solvable using the analytical methods, such that the fracture toughness could be calculated in the closed form. The semi-infinite crack is parallel to the direction of one, say, X-fiber. The crack edges are subjected to action of crack opening forces, such that at infinity) crack opening exponentially decrease.

Consider now the material containing the semi-infinite crack (**Fig.3**). Semi-infinite crack is local separation of material with infinite length, with crack edges which are extremely close together and with extremely sharp crack tip. This means, that all the Y-fibers with the number, less than zero are broken between the zeroth and the first X-fiber. Let the both edges of the crack are subjected to opening loads. The loads, applied at infinity, can also be taken into account making use of superposition principle. The loads on the both edges

$$f_n, \quad n < 0.$$

The loads are applied to the broken ends of the same fiber, have the equal magnitudes, but the opposite direction. In order to apply the Eq. (1.3), valid for perfect grid without damage of fibers, to the problem with semi-infinite crack, superposition principle is used. According to this principle, we consider auxiliary perfect grid without crack, but with the same deformation and stress distribution and subjected to the action of the same external forces, as the cracked grid. To provide the identity of stress and deformation state of this two grids everywhere (except. naturally, crack line) it is necessary to add the certain additional forces. The additional forces are applied to the fibers in auxiliary perfect grid which are corresponding to the ruptured fibers of cracked layer. The auxiliary perfect grid with the external forces distribution

$$q_{nk} = \begin{cases} \dfrac{\beta}{\Delta}(v_{n0} - v_{n1}) + f_n & \text{for} \quad n < 0, \quad k = 0, \\ -\dfrac{\beta}{\Delta}(v_{n0} - v_{n1}) - f_n & \text{for} \quad n < 0, \quad k = 1, \\ 0 & \text{for} \quad n \geq 0, \end{cases}$$

has the same stress-deformation state, as layer with forces $f_n$, applied to the edges of the crack.

Due to the symmetry

$$q_{n0} = -q_{n1},$$
$$v_{n0} = -v_{n1},$$
$$q_{n0} = \dfrac{2\beta}{\Delta} v_{n0} + f_n, \quad \text{for} \quad n < 0.$$

The coefficients $q_{nk}$ generate the corresponding Fourier series:

(3.1) $\quad Q(z,w) = \sum\limits_{n=-\infty}^{\infty} \sum\limits_{k=-\infty}^{\infty} q_{nk} z^n w^k = \left( \dfrac{2\beta}{\Delta} W_-(z) + F_-(z) \right)(1-w),$

(3.2) $\quad V(z,0) \equiv W(z) = W_+(z) + W_-(z).$

In the equations (3.1) and (3.2) the expressions

$$W_-(z) = \sum\limits_{n=-\infty}^{-1} v_{n0} z^n,$$

$$F_-(z) = \sum\limits_{n=-\infty}^{-1} f_n z^n,$$

are the principal parts of series (with the subscript "minus"). The regular part (with the subscript "plus") is

$$W_+(z) = \sum\limits_{n=0}^{\infty} v_{n0} z^n.$$

## 3.2 Reduction to Riemann-Hilbert problem

Substitution (3.1) into (1.3), division by w and integration over the unit circle $|w| = 1$:

$$V(z,0) = \frac{1}{2\pi i} \oint_{|w|=1} \frac{V(z,w)dw}{w} =$$

(3.3)

$$= W_+(z) + W_-(z) = \left[ W_-(z) + \frac{\Delta}{2\beta} F_-(z) \right] H_\lambda(z).$$

Using the expression (2.15) for $T(w)$ the integral (3.3) is

$$H_\lambda(z) = \frac{\beta}{\pi i \Delta} \oint_{|w|=1} \frac{(1-w)dw}{\left[ \frac{1}{T(w)} + \frac{1}{B_\lambda(z)} \right] w} =$$

$$= \frac{\beta}{\pi i \Delta} \oint_{|w|=1} \frac{(1-w)dw}{\left[ \frac{\beta}{\Delta}(1-w)\left(1 - \frac{1}{w}\right) + \frac{1}{B_\lambda(z)} \right] w}.$$

Integral evaluates in the closed form

$$H_\lambda(z) = 1 - \frac{1}{\sqrt{1 + \frac{4\beta}{\Delta} B_\lambda(z)}}.$$

The $B(z)$ is given by the formula (2.9) for Bernoulli-Euler beam. The kernel function is

(3.4) $$L_\lambda(z) \equiv 1 - H_\lambda(z) = \frac{1}{\sqrt{1 + \frac{4\beta}{\Delta} B_\lambda(z)}}.$$

Substitution (3.2) into (3.3) leads to the following Riemann-Hilbert problem [20]

(3.5) $$W_+(z) + L(z)W_-(z) = \frac{1}{2\beta} H_\lambda(z) F_-(z).$$

The solution of (3.5) gives the complete description of semi-infinite crack problem.

The number of zeros in kernel function $L_\lambda(z)$ along the path $|z| = 1$ changes the index of problem (3.5). Correspondingly, the number of homogeneous solutions of problem (3.5) increases. This can be interpreted as appearance of instability of layer under compression. Namely, the equation

$$L_\lambda(z) = 0$$

has the spectrum of real roots for all $\lambda < 0$. This means that the media cannot resists the compressing forces. Case $\lambda = 0$ is critical and corresponds to the changing of equation type and instability.

## 4  Semi-infinite crack in the presence of tension in orthogonal

# fibers ($\lambda > 0$)

## 4.1 The factorization method for solution

The solution for crack in the presence of tension in the family of fibers that are parallel to the crack edges is explained in this section. The parameter that characterizes the pre-stress is positive $\lambda > 0$.

For the solution of Riemann-Hilbert problem the Wiener-Hopf method is applied [21, 22].

Function $L_\lambda(z)$ vanishes at the point $|z|=1$. To apply the factorization method for this problem the new function is introduced

(4.1) $$\Phi(z) = \frac{L_\lambda(z)}{\sqrt{(1-z)(1-1/z)}}.$$

The function $\Phi(z)$ is a positive real function on the unit circle. Consequently the index of it is equal to zero.

Factorization of the function $\Phi(z)$ is the representation as a product of two functions

$$\Phi(z) = \Phi_-(z)\Phi_+(z).$$

Factorization of this function $\Phi(z)$ is performed according the formulae [23]

(4.2) $$\Phi_+(z) = \exp\left(\frac{1}{2\pi i}\oint_{|w|=1} \frac{\ln \Phi(w)dw}{w-z}\right) \qquad \text{for} \qquad |z|<1,$$

$$\Phi_-(z) = \exp\left(-\frac{1}{2\pi i}\oint_{|w|=1} \frac{\ln \Phi(w)dw}{w-z}\right) \qquad \text{for} \qquad |z|>1.$$

Function $\Phi_+(z)$ a holomorphic in circular region $|z|<1$, and $\Phi_-(z)$ is a holomorphic function in a region $|z|>1$.

Rewrite the equation (3.5) in the form

(4.3) $$\frac{W_+(z)}{\sqrt{1-z}\,\Phi_+(z)} - \chi_-(z) = -\sqrt{1-z^{-1}}\,\Phi_-(z)W_-(z) + \chi_+(z),$$

$$\chi(z) = \frac{\Delta}{2\beta} \frac{H_\lambda(z)\,F_-(z)}{\sqrt{1-z}\,\Phi_+(z)},$$

$$\chi_-(w) = \frac{\Delta}{2\beta} \frac{1}{2\pi i} \oint_{|z|=1} \frac{\sqrt{1-z}\, F_-(z) H_\lambda(z) dz}{\Phi_+(z)(z-w)} \qquad \text{for} \qquad |w| > 1,$$

$$\chi_+(w) = \frac{\Delta}{2\beta} \frac{1}{2\pi i} \oint_{|z|=1} \frac{\sqrt{1-z}\, F_-(z) H_\lambda(z) dz}{\Phi_+(z)(w-z)} \qquad \text{for} \qquad |w| < 1.$$

The first term in the left side of (4.3) is a holomorphic function in the region $|z| < 1$, the second-holomorphic in the infinite region $|z| > 1$.

From (4.3) follows that the solution of non-homogeneous Riemann-Hilbert problem, vanishing at infinity is

(4.4) $\qquad W_-(z) = \dfrac{\chi_-(z)}{\sqrt{1-z^{-1}}\, \Phi_-(z)},$

$\qquad\qquad W_+(z) = \chi_+(z) \Phi_+(z) \sqrt{1-z}.$

With the expression (4.4) the stress in the first unbroken fiber reads

(4.5)
$$p_0 = \frac{\beta}{\Delta}(v_{10} - v_{00}) = \frac{2\beta}{\Delta} u_{00} =$$
$$= \frac{2\beta}{\Delta} W_+(0) = \frac{\Phi_+(0)}{2\pi i} \oint_{|z|=1} \frac{F_-(z) H_\lambda(z) dz}{\Phi_+(z) z \sqrt{1-z}}.$$

## 4.2 The asymptotic behavior of the Fourier coefficients

Our nearest aim is to study the asymptotic behavior of the Fourier coefficients of the functions $W_-(w)$ and $W_+(w)$. These functions determine the behavior of stress and displacement in fibers remote from crack origin. It could be demonstrated that the direct estimation of coefficients of expressions (4.4)-(4.5) delivers no useful information. The explanation of this problem is the following. The Fourier coefficients of the functions $W_-(w)$ and $W_+(w)$ are exponentially decrease at infinity for the exponentially decreasing load $F_-(z)$. This asymptotic behavior can be called long-scale asymptotic. On the contrary, we are interested in estimation an intermediate asymptotic behavior. Solely with the expressions for intermediate asymptotic behavior it is possible to calculate the stresses of fibers on the crack tip.

We evaluate the intermediate asymptotic behavior with the method of exponentially decreasing load sequence is applied. The exponentially distributed load on the row of fibers is considered:

(4.6) $\quad f_n = N\tau^{-n}, \quad 0 < \tau < 1, \quad N > 0, \quad \text{for} \quad n = -1,-2,-3,\ldots$

$$F_-(z) = \sum_{n=-\infty}^{-1} f_n z^n = N \sum_{n=-\infty}^{-1} \left(\frac{z}{\tau}\right)^n = N \sum_{n=1}^{\infty} \left(\frac{\tau}{z}\right)^n = N \frac{\tau}{z-\tau}.$$

The value $N$ characterizes the load on the fibers. In the limit case $\tau \to 1$ all fibers on the row will be pressed with the constant load normal to crack edges. Unfortunately, the calculations $\tau = 1$ could not be completed due to singularity. We estimate the asymptotes of the expressions (4.6) with the term

$$1/\sqrt{1-\tau}.$$

Integrals (4.4)-(4.5) for the load (4.6) could be expressed in closed form. For the case $\lambda > 0$ we obtain

(4.7) $\quad W_-(w) = \dfrac{N}{2\beta} \dfrac{\tau H_\lambda(\tau)}{\Phi_-(w)\Phi_+(\tau)(w-\tau)\sqrt{1-\tau}\sqrt{1-w^{-1}}} \quad$ for $\quad |w| > 1,$

(4.8) $\quad W_+(w) = \dfrac{N}{2\beta} \dfrac{\Phi_+(w)\sqrt{1-w}}{\dfrac{w}{\tau}-1} \left[ \dfrac{H_\lambda(\tau)}{\Phi_+(\tau)\sqrt{1-\tau}} - \dfrac{H_\lambda(w)}{\Phi_+(w)\sqrt{1-w}} \right] \quad$ for $\quad |w| < 1.$

Consider the asymptotic behavior of (4.7)-(4.8) at the limit case $\tau \to 1$. If $\tau \to 1$, the external loads tends to a constant load on each fiber in the row. The first task is to determine the main term of the asymptotical expansion in (4.8) with respect to $\sqrt{1-\tau}$.

Finally we have to estimate the asymptotical behavior of the Fourier coefficient of the obtained main term at infinity. This estimation gives the desired intermediate asymptotic behavior.

## *4.3 The opening of crack for the constant load*

From (4.7)-(4.8) follow that the main terms of asymptotic expansion with respect to $\sqrt{1-\tau}$ are

(4.9) $\quad W_-(w) \underset{\tau \to 1}{\sim} W_-^*(w) = \dfrac{N\Delta}{2\beta} \dfrac{1}{\Phi_-(w)\Phi_+(1)(w-1)\sqrt{1-\tau}\sqrt{1-w^{-1}}} \quad$ for $\quad |w| > 1,$

(4.10) $\quad W_+(w) \underset{\tau \to 1}{\sim} W_+^*(w) = \dfrac{N\Delta}{2\beta} \dfrac{\Phi_+(w)}{\Phi_+(1)\sqrt{1-\tau}\sqrt{1-w}} \quad$ for $\quad |w| < 1,$

while

$$H_\lambda(\tau) \to 1 \qquad \text{as} \qquad \tau \to 1.$$

Consider now the asymptotic behavior of the coefficients $v_{n0}$. These coefficients describe the opening of the nodal points in the first row. The behavior of the coefficients $v_{n0}$ is governed by the power series expansions $W_-^*(w)$ and $W_+^*(w)$ at infinity. The Cauchy formula reads:

$$(4.11) \qquad v_{n0} = \begin{cases} \dfrac{1}{2\pi i} \oint_\Gamma \dfrac{W_+^*(w)dw}{w^{n+1}} & \text{for } n > 0, \\[2mm] \dfrac{1}{2\pi i} \oint_\Gamma \dfrac{W_-^*(w)dw}{w^{n+1}} & \text{for } n < 0. \end{cases}$$

At first consider the integral with the function $W_+^*(w)$. Move the contour $\Gamma$ away from the origin as far as possible, and turn it to a contour composed of two straight lines along the both edges of the branch cut, radiating away from the origin. For contour $\Gamma$ we take the upper and lower edges of the cut at the positive real axis from 1 to $\infty$.

Putting $z = \exp \chi$ we obtain

$$v_{n0} = \frac{\exp(\pi i/2) - \exp(-\pi i/2)}{2\pi i} \frac{N\Delta}{2\beta \Phi_+(1)\sqrt{1-\tau}} \int_0^\infty \frac{\Phi_+(\exp \chi)\exp(-n\chi)}{\sqrt{1-\exp \chi}} d\chi \qquad \text{for} \qquad n \to \infty.$$

The theorem about the asymptotic behavior of the Laplace transformation [24,25] is used. With this theorem the desired asymptotic estimations for the integral for $v_{n0}$ and $p_n$ are:

$$(4.12) \qquad v_{n0} \sim \frac{N\Delta}{2\beta\sqrt{\pi n(1-\tau)}}, \qquad p_n \sim \frac{N}{\sqrt{\pi n(1-\tau)}} \qquad \text{for} \qquad n \to \infty.$$

The formulas (4.12)-(4.13) describe the displacements and loads in fibers on the unbroken side with respect to the crack tip. The fibers with the positive numbers $n$ connect the eventual crack edges together. The deflections and fiber loads gradually disappear with the increasing distance from the tip as

$1/\sqrt{n}$ with $n \to \infty$.

The asymptotic behavior of Fourier coefficients for $v_{n0}$ with negative numbers $n$ as is studied similarly. In this case the asymptotic behavior of $v_{n0}$ (4.11) as $n \to -\infty$ is studied. The contour $\Gamma$ in this case is chosen as upper and lower edges of the cut $0 \leq z \leq 1$. The asymptotic estimation for the integral for $v_{n0}$ is given by

$$(4.13) \quad v_{n0} \sim \frac{2N\Delta\sqrt{|n|}}{2\beta\sqrt{\pi(1-\tau)}} \quad \text{for} \quad n \to -\infty.$$

This formula describes the displacements on the opened side of the crack tip, where the fibers are broken and do not connect the edges of the crack any more. From (4.13) follows that the crack opening increases as $\sqrt{|n|}$ with $n \to -\infty$.

## 4.4 Stresses in the first fiber on the crack tip

The calculation of load in the first fiber on the crack tip the equation (4.5) is applied. The substitution (4.6) in (4.5) delivers

$$(4.14) \quad p_0 = \frac{N}{\sqrt{1-\tau}} \frac{\Phi_+(0)}{\Phi_+(1)} + o\left(\frac{1}{\sqrt{1-\tau}}\right) \quad \text{for} \quad \tau \to 1.$$

Comparison of (4.13) and (4.14) shows, that the stresses in fibers and opening the crack can be characterized by means of coefficient

$$K = \frac{\sqrt{2}}{\sqrt{\pi\Delta(1-\tau)}} N.$$

This coefficient plays the role of the stress intensity factor as known from elasticity theory [26]. With this coefficient the opening of crack and the stresses in fibers read

$$(4.15) \quad v_{n0} \sim \frac{K}{\sqrt{2}\beta}\sqrt{r}, \quad \text{for} \quad n \to -\infty.$$

In (4.15) the symbol

$$r = n\Delta$$

designates the distance from the crack tip.

The loads in the fibers read

$$(4.16) \quad p_0 = \sqrt{\frac{\pi\Delta}{2}} \frac{\Phi_+(0)}{\Phi_+(1)} K,$$

(4.17) $\quad p_n = \dfrac{2\beta}{\Delta} v_{n0} \sim \dfrac{\Delta}{\sqrt{2r}} K$.

The formula (4.16) comprises the dimensionless coefficient

$$\Phi_+(1)/\Phi_+(0).$$

This coefficient expresses the influence of material parameters on the stress in the first unbroken fiber on the crack tip.

Using the relations (4.2) the closed form solution for this coefficient is obtained

(4.18) $\quad \ln \Phi(0) = \dfrac{1}{2\pi} \int\limits_{-\pi}^{\pi} \varphi(\xi)d\xi$,

$$\ln \Phi(1) = \dfrac{1}{2}\ln \Phi(0) + \dfrac{1}{2}\varphi(0),$$

where

$$\varphi(\xi) = \Phi(e^{i\xi}) = \dfrac{1}{2}\ln\left[\sinh^2\left(\dfrac{\eta}{2}\right) + \sinh^2\left(\dfrac{\xi}{2}\right)\right] - \dfrac{1}{2}\ln\left[\sin^4\left(\dfrac{\xi}{2}\right) + 2A_\lambda \sin^2\left(\dfrac{\xi}{2}\right) + B_\lambda\right].$$

For Bernoulli-Euler beam

$$2A_\lambda = \sinh\left(\dfrac{\eta}{2}\right)\left[\sinh\left(\dfrac{\eta}{2}\right) - \dfrac{2\alpha^{1/2}\beta}{\lambda^{3/2}\Delta}\cosh\left(\dfrac{\eta}{2}\right)\right] + \dfrac{\beta}{\lambda},$$

$$B_\lambda = \dfrac{\beta}{\lambda}\sinh^2\left(\dfrac{\eta}{2}\right),$$

$$C_\lambda = 1/\sinh^2\left(\dfrac{\eta}{2}\right).$$

For Timoshenko beam

$$2A_\lambda = \sinh\left(\dfrac{\eta}{2}\right)\left[\sinh\left(\dfrac{\eta}{2}\right) + \dfrac{2(\lambda-\gamma)\alpha^{1/2}\beta}{\lambda^{3/2}\gamma\Delta}\cosh\left(\dfrac{\eta}{2}\right)\right] + \dfrac{\beta}{\lambda},$$

$$B_\lambda = \dfrac{\beta}{\lambda}\sinh^2\left(\dfrac{\eta}{2}\right),$$

$$C_\lambda = 1/\sinh^2\left(\frac{\eta}{2}\right).$$

Application of the identity:

$$\frac{1}{2\pi}\int_0^{2\pi} \ln(1+x\sin^2 t)dt = \ln\frac{1+\sqrt{1+x}}{2}$$

provides the closed form solution for the case $\lambda > 0$

(4.19) $$\frac{\Phi_+(0)}{\Phi_+(1)} = \frac{2\left(1+\sqrt{1+C_\lambda}\right)}{1+\sqrt{1+(2A_\lambda+1)/B_\lambda}+\sqrt{2}\sqrt{1+A_\lambda/B_\lambda}+\sqrt{1+(2A_\lambda+1)/B_\lambda}},$$

(4.20) $$\lim_{\lambda\to 0}\frac{\Phi_+(0)}{\Phi_+(1)} = 1.$$

# 5  Semi-infinite crack in the absence of tension in orthogonal fibers ($\lambda = 0$).

## *5.1 Factorization in the absence of tension*

The solution in the absence of tension in the family of fibers that are parallel to the crack edges is considered in this section. The parameter that characterizes the pre-stress is zero $\lambda = 0$.

The new function $\Psi(z)$, which assumes only the real positive values on the unit circle $|z|=1$ is introduced

(5.1) $$\Psi(z) = L_0(z)(1-z)\left(1-\frac{1}{z}\right).$$

Factorization of function $\Psi(z)$ delivers the expressions, similar to (4.2). Riemann-Hilbert problem assumes the form

(5.2) $$\frac{W_+(z)}{(1-z)\Psi_+(z)} + \left(1-\frac{1}{z}\right)\Psi_-(z)W_-(z) = \frac{\Delta}{2\beta}\frac{H_0(1)F_-(z)}{(1-z)\Psi_+(z)}.$$

Solution of (5.2) is analogous to (4.4). For the load (4.6) the solution reads:

(5.3) $$W_-(w) = \frac{N\Delta}{2\beta}\frac{\tau H_0(\tau)}{\Psi_-(w)\Psi_+(\tau)(w-\tau)(1-\tau)(1-1/w)} \qquad \text{for} \qquad |w|>1,$$

$$W_+(w) = \frac{N\Delta}{2\beta} \frac{1-w}{1-w/\tau} \left[ \frac{H_0(\tau)}{(1-\tau)\Psi_+(\tau)} - \frac{H_0(w)}{(1-w)\Psi_+(w)} \right] \Psi_+(w) \qquad \text{for} \qquad |w| < 1.$$

The intermediate asymptotic expansions are generated by the main terms of asymptotical expansion with respect to $1-\tau$:

(5.4) $\quad W_-(w) \underset{\tau \to 1}{\sim} W_-^{**}(w) = \dfrac{N\Delta}{2\beta} \dfrac{1}{\Psi_-(w)\Psi_+(1)(w-1)(1-\tau)(1-1/w)} \qquad$ for $\qquad |w| > 1$,

$\quad W_+(w) \underset{\tau \to 1}{\sim} W_+^{**}(w) = -\dfrac{N\Delta}{2\beta} \dfrac{\Psi_+(w)}{\Psi_+(1)(1-\tau)(1-w)} \qquad$ for $\qquad |w| < 1.$

## 5.2 Stresses in the first unbroken fiber on the crack tip

Fourier coefficients of function $W_-^{**}(w)$ have the following asymptotic behavior

(5.5) $\quad v_{n0} \sim \dfrac{N|n|\Delta}{2\beta} \dfrac{1}{1-\tau} \qquad$ for $n \to -\infty$.

The stress in the first unbroken fiber is equal to

(5.6) $\quad p_0 = \dfrac{\Psi_+(0)}{\Psi_+(1)} \dfrac{N}{1-\tau}.$

The expression (5.5) demonstrates, that in the absence of the tension, parallel to the crack direction, edges of the crack form the finite angle at the tip

$$\Lambda = \arctan \frac{N}{2\beta(1-\tau)}.$$

In contrast to the case $\lambda > 0$, the shape of the crack is lens-shaped with the sharp angle on the tip.

The stress in the first fiber (5.6) is proportional to tangent $\Lambda$:

(5.7) $\quad p_0 = 2\beta \dfrac{\Psi_+(0)}{\Psi_+(1)} \tan \Lambda.$

Comparison on (4.13) and (5.5) demonstrates the difference of the asymptotical behavior in this two cases; the former contains factor

$$\frac{1}{\sqrt{1-\tau}}$$

while the latter - the factor

$$\frac{1}{1-\tau}.$$

This means that the first unbroken fiber in the case $\lambda = 0$ carries much more load, then when the tension exists $\lambda > 0$. The crack becomes instable and spreads at lover load.

The physical explanation of this behavior is direct. The tension of fibers ($\lambda > 0$), which are parallel to crack edges, makes these fibers stiffer. The load in the unbroken fibers, which are normal to the crack edges, evenly redistributes over the flawless fibers behind the crack tip. Alternatively If the tension vanishes ($\lambda = 0$) only immediately adjacent to the crack tip faultless fibers carry the additional load. The finite-element model of the double-periodic lattices confirms this behavior. On the **Fig.4** the results of finite-element simulation of the long crack loaded on its boundaries depicted. The **Fig. 5** shows the local break in the finite-element mesh of the long elastic fibers, loaded by constant forces.

# 6 Continuous equations and fracture of the finite length crack

## *6.1 Continuous equations*

In this section it is studied the macroscopic equations of the material which are obtained as a limit equations when the character size of material structure turns to zero. These equations describe the averaged continuous medium and valid remote of crack tip zone. The usage of the macroscopic equation enables to estimate stress intensity factor at the crack tip and check the conditions of crack growth. These estimations perform with the aid of intermediate asymptotic expansions. Namely suppose that a solution of macroscopic boundary problem is obtained. Asymptotical estimations of opening near the crack tip give the coefficients of intermediate asymptotic expansions and usage of formulae (4.16) and (5.7). This procedure enables to evaluate the stress in the first unbroken fiber and so check the fracture conditions.

Our first goal is obtaining of the continuous macroscopic equation from the discrete one. For this purpose multiply all linear dimensions (i.e. diameters of fibers and distances between the fibers) on a small parameter $\varepsilon$ and in the obtained equations turn it to zero. The macroscopic equations are obtained at as a limit when $\varepsilon \to 0$. The most effective method for this seems to be calculation of the limit of Fourier trans-formations of discrete equations.

Fourier transforms of the differential operators of continuous macroscopic equations are the limit case of Fourier transforms of the finite-difference operators of discrete equations. As the distance between fibers $\varepsilon\Delta$ vanishes, the transforms for operators appear as the limits

(6.1) $\quad T^*(\xi) = \dfrac{1}{\sqrt{2\pi}} \lim_{\varepsilon \to 0} T(e^{\varepsilon i \Delta \xi})$,

$\qquad B_\lambda^*(\xi) = \dfrac{1}{\sqrt{2\pi}} \lim_{\varepsilon \to 0} B_\lambda(e^{\varepsilon i \Delta \xi})$,

$\qquad B_\gamma^*(\xi) = \dfrac{1}{\sqrt{2\pi}} \lim_{\varepsilon \to 0} B_\gamma(e^{\varepsilon i \Delta \xi})$.

Inversion of the Fourier transforms gives the continuous macroscopic equations of media in linear approximation:

(6.2) $\quad \left( \dfrac{\partial^2}{\partial x^2} - \dfrac{\alpha}{\beta} \dfrac{\partial^4}{\partial y^4} \right) u(x, y) = p(x, y) \qquad$ for $\quad \gamma = 0$,

(6.3) $\quad \left( \dfrac{\partial^2}{\partial x^2} + \dfrac{\gamma}{\beta} \dfrac{\partial^2}{\partial y^2} \right) u(x, y) = p(x, y) \qquad$ for $\quad \gamma > 0$,

(6.4) $\quad \left( \dfrac{\partial^2}{\partial y^2} - \dfrac{\alpha}{\beta} \dfrac{\partial^4}{\partial x^4} \right) v(x, y) = q(x, y), \qquad$ for $\quad \lambda = 0$,

(6.5) $\quad \left( \dfrac{\partial^2}{\partial y^2} + \dfrac{\lambda}{\beta} \dfrac{\partial^2}{\partial x^2} \right) v(x, y) = q(x, y), \qquad$ for $\quad \lambda > 0$,

with

$\qquad u(x, y)$, $v(x, y)$ are the components of macroscopic displacements

$\qquad p(x, y)$, $q(x, y)$ are the components of surface load.

## 6.2 Finite length crack in continuous material

Consider the case of finite length crack. Suppose that fibers are ruptured along the interval $-1 \le x \le 1$ of the axis $y = 0$. The material at infinity is subjected to the action of opening stress with the magnitude $f(x)$. By superposition principle, the same stress distribution occurs if the loads $f(x)$ are on the edges of the cut. The stress-deformation state near the crack is described by the following boundary value problem for the equations (6.3) and (6.4):

(6.6) $\quad v(x,0) = 0 \quad$ for $\quad |x| > 1$,

$$-\frac{\partial v}{\partial y}(x,0) = Q(x) \quad \text{for} \quad |x| < 1,$$

where

$$Q(x) = f(x)\Delta / \beta$$

and

$$q(x, y) = 0.$$

The solution of the boundary value problem (6.6) for the equation

(6.7) $\quad L^M v(x, y) = 0$

with

(6.8) $\quad L^M \equiv \dfrac{\partial^2}{\partial y^2} - \omega^2 \left( i \dfrac{\partial}{\partial x} \right)^{2M}$

is obtained with the aid of Fourier transforms method [27]. The solution of boundary value problem (6.6)-(6.7) is given by the integral with the Bessel functions

(6.9) $\quad v(x, y) = \dfrac{1}{\pi} \int\limits_{-\infty}^{\infty} \xi^{-1/2} \exp\left(- y\omega\xi^M - i\xi x\right) I(\xi) d\xi,$

(6.10) $\quad I(\xi) = \dfrac{(2x)^{1-\frac{M}{2}}}{\omega \Gamma(M/2)} \sqrt{\dfrac{\pi}{2}} \int\limits_0^1 \eta^{\frac{1+M}{2}} J_{\frac{M-1}{2}}(\eta x) \int\limits_0^1 Q(\zeta\eta)(1-\zeta^2)^{\frac{M}{2}-1} d\zeta d\eta.$

The formulas (6.9)-(6.1) deliver the solutions of boundary value problem with different values of parameters $M$ and $\omega$. Namely, the solution of the equation (6.3) results from (6.9)-(6.10) if the parameters are chosen as

$$M = 2,$$

$$\omega^2 = \frac{\alpha}{\beta}.$$

The formulas (6.9)-(6.10) deliver solution of the equation (6.4) with

$$M = 1,$$

$$\omega^2 = \frac{\lambda}{\beta}.$$

Particularly, there exists the closed form of integrals (6.9)-(6.10) for $Q(x)=1$:

$$\begin{aligned}
(6.11) \quad v(x) = v(x,0) &= \frac{2^{\frac{M-1}{2}}}{\omega\Gamma\left(\frac{M+1}{2}\right)} \int_0^1 \eta^{\frac{1+M}{2}} \int_0^\infty \cos(\xi x) \xi^{\frac{M-1}{2}} J_{\frac{M-1}{2}}(\eta\xi) d\xi d\eta = \\
&= \begin{cases} \omega^{-1}\sqrt{1-x^2} & \text{for} \quad M=1 \text{ or } \quad \lambda > 0, \\ \frac{\sqrt{\pi}}{2^{3/2}\omega}(1-x^2) & \text{for} \quad M=2 \text{ or } \quad \lambda = 0. \end{cases}
\end{aligned}$$

The stress in the first fiber on the tip of crack could be calculated immediately using the formulae (4.14), (5.7).

The stress intensity factor can be defined in terms of crack opening using the expression (4.16):

$$(6.12) \quad K = \beta\sqrt{2} \lim_{r \to +0} \frac{v(1-r)}{\sqrt{r}} = \beta\sqrt{2} \lim_{x \to 1-} \frac{v(x,0)}{\sqrt{1-x}}.$$

Calculating the corresponding limit in (6.12) one can obtain the closed form expression for stress intensity factor.

The stress intensity factor depends upon fiber tension and material parameters. Remarkably, that in the presence of stabilized tension the macroscopic behavior of the material is similar to the behavior of homogeneous elastic medium. In this case the crack in material assumes the elliptic shape [28]. Both the ellipse curvature at the tip and coefficient of the stress singularity are linearly dependent upon the value of stress intensity factor $K$. In its turn, the stress intensity factor $K$ is defined by external load distribution according to the formulas (6.6) and (6.7).

Our next aim is to derive the fracture condition in terms of stress intensity factor $K$. For this purpose the equation (4.16) is used. This expression demonstrates that the tension in the first unbroken fiber $p_0$ depends linearly upon stress intensity factor $K$.

Fracture of the first fiber occurs, when this tension exceeds its limit value $p_{max}$.

In other words, the crack growth begins, when stress intensity factor reaches the critical value

(6.13) $\quad K^* = p_{max} \sqrt{\dfrac{2}{\pi \Delta} \dfrac{\Phi_+(1)}{\Phi_+(0)}}$ .

With the formula (4.19) the expression (6.13) delivers the critical stress intensity factor $K^*$ as the function of material parameters.

# 7 Conclusions

This article present the double-periodical lattice made of infinite elastic fibers that withstand bending and tension. The model describes the elastic properties of flat periodic structure. With this model the behavior of a two-dimensional array of infinite fibers is simulated.

The material that contains a row of broken fibers is considered. These broken fibers form the failure in the material that shapes like a long straight crack. The lattice is tensioned in the direction, which is orthogonal to the direction of straight crack. The conditions of fracture of this lattice are investigated. The closed form expression for the stress in the first unbroken fiber and the expression for fracture toughness are given. These values are the functions of mechanical parameters of lattice and tensions in both families of fibers.

The closed form solution demonstrates a notable behavior of the material. Namely, the fracture behavior of two-dimensional lattice is cardinally depends upon the pre-stress in the material in the direction, parallel to crack direction.

If the tension in fibers that parallel to the crack direction exists, it stabilizes the crack growth and makes the load distribution in the unbroken fibers more even. The two-dimensional lattice behaves in the presence of tension in both directions similarly to the plane elastic media. The finite length crack assumes the shape of the elongated elliptic split. Crack opening in the presence of tension in both directions is proportional to

$$\sqrt{l^2 - x^2} ,$$

where x- the distance from crack center, $l$ - crack length.

Another behavior of lattice occurs if the fibers, parallel to crack direction, are unstressed. The character of stress concentration near the crack differs. The load distribution at the crack tip varies considerably. The first unbroken fiber carries higher load. The crack assumes the parabolic shape, i.e. opening is proportional to

$$l^2 - x^2 .$$

The crack is lens-shaped and the crack borders form at the tip the finite angle.

**References**

[1] Ashby M.F. (1983) The mechanical properties of cellular solids. Metall. Trans A 14:1755–1769

[2] Gibson L.J., Ashby M.F. (1997) Cellular solids. Structure and properties. Cambridge University Press

[3] Quintana-Alonso I., Fleck N. A. (2009) Fracture of Brittle Lattice Materials, A Review, in: I.M. Daniel et al. (eds.), Major Accomplishments in Composite Materials and Sandwich Structures: An Anthology of ONR Sponsored Research, Springer Science+Business Media B.V.

[4] Ostoja-Starzewski M. (2002) Lattice models in micromechanics, Appl. Mech. Rev. 55, 1, 35-60

[5] Lipperman F., Ryvkin M., Fuchs M. B. (2007) Fracture toughness of two-dimensional cellular material with periodic microstructure, Int J Fract 146:279–290, DOI 10.1007/s10704-007-9171-5

[6] Rinaldi A., Krajcinovic D., Peralta P., Lai Y.-C. (2008) Lattice models of polycrystalline microstructures: A quantitative approach, Mechanics of Materials, 40, 17–36

[7] Romijn N. E.R., Fleck N. A. (2007) The fracture toughness of planar lattices: Imperfection sensitivity, Journal of the Mechanics and Physics of Solids, 55, 2538–2564

[8] Zhou S. J., Carlsson A. E., Thomson R. (1993) Dislocation nucleation and crack stability: Lattice Green's-function treatment of cracks in a model hexagonal lattice, PHYS. REV. B, 47, 13

[9] Tserpes K. I., Silvestre N. (eds.) (2014) Modeling of Carbon Nanotubes, Graphene and their Composites, Springer Series in Materials Science, 188

[10] Savage G., (1993) Carbon-Carbon Composites, Springer Netherlands

[11] Fitzer E., Manocha L. M. (1998) Carbon Reinforcements and Carbon/Carbon Composites, Springer Berlin Heidelberg

[12] Morgan P. (2005) Carbon Fibers and their Composites, CRC Press, Taylor & Francis Group, Boca Raton

[13] Hedgepeth, J. M. Van Dyke, P. (1967) Local Stress Concentrations in Imperfect Filamentary Composite Materials," Journal of Composite Materials, 1, 294-309

[14] Van Dyke, P. Hedgepeth, J. M. (1969) Stress Concentrations from Single-Filament Failures in Composite Materials, Textile Research Journal, 39, 618-626

[15] Hikami F., Chout T.-W. (1990) Explicit Crack Problem Solutions of Unidirectional Composites: Elastic Stress

Concentrations, AIAA JOURNAL, 28, 3,499

16 Liu A. F., (2005) Mechanics and Mechanisms of Fracture: An Introduction, ASM International, Materials Park, Ohio 44073-0002, www.asminternational.org

[17] Henrici P. (1986) Applied and Computational Complex Analysis. Vol. I - III. Wiley. New York

[18] Schwartz L. (1973 ) Theorie des Distributions. Hermann, Paris

[19] Gradshtein I.S., Ryzhik I.S. (1965) Tables of Integrals, Series and Products. Academic Press Inc., New York

[20] Bitsadze, A.V. (2001) Boundary value problems of analytic function theory, in Hazewinkel, Michiel, Encyclopedia of Mathematics, Springer, ISBN 978-1-55608-010-4

[21] Paley R, Wiener N. (2012) Fourier Transforms in the Complex Domain, Colloquium Publications, American Mathematical Society, Band 19, Literary Licensing, LLC

[22] Noble B. (1958) Methods Based on the Wiener-Hopf Technique for the Solution of Partial Differential Equations, Chelsea Publishing Company

[23] Khimshiashvili, G. (2001) Birkhoff factorization, in Hazewinkel, Michiel, Encyclopedia of Mathematics, Springer, ISBN 978-1-55608-010-4

[24] De Bruijn N. G. (1970) Asymptotic Methods in Analysis. North Holland, Amsterdam

[25] Lauweder H.A. (1977) Asymptotical Analysis. Math. Centrum, Amsterdam

[26] Sneddon, I. N. (1946) The distribution of stress in the neighbourhood of a crack in an elastic solid, Proceedings of the Royal Society A 187 (1009): 229

[27] Sneddon I.N. (1951) Fourier Transforms. McGraw-Hill, New York

[28] Sih G.C., Liebowitz H. (1968) Mathematical Theories of Brittle Fracture. In: Fracture. Vol.2. Academic Press. New York